# Investigating the Impact of Learners' Learning Styles on the Acceptance of Open Learner Models for Information Sharing

**YongWee Sek**[1, 2]
[1] School of Business IT and Logistics
RMIT University
Melbourne, Victoria, Australia
Email: yongwee.sek@rmit.edu.au

[2]Faculty of Information and Communication Technology
Universiti Teknikal Malaysia Melaka
Melaka, Malaysia
Email: ywsek@utem.edu.my

**Hepu Deng**
School of Business IT and Logistics
RMIT University
Melbourne, Victoria, Australia
Email: hepu.deng@rmit.edu.au

**Elspeth McKay**
School of Business IT and Logistics
RMIT University
Melbourne, Victoria, Australia
Email: elspeth.mckay@rmit.edu.au

**Minghui Qian**
Smart City Research Center
School of Information Resources Management
Renmin University of China
Beijing, China
Email: minghuiqian@163.com

## Abstract

Individual differences in learners' learning styles can have a significant effect on their acceptance of collaboration technologies to facilitate the sharing of learning information in technology-based collaborative learning. There is, however, a lack of understanding of the impact of learning styles on the acceptance of open learner models as a collaboration technology for information sharing. This study investigates the impact of learners' learning styles on their acceptance of open learner models for information sharing. A total of 240 undergraduate students in a university in Malaysia have participated in the online survey. A chi-square test is performed to explore the relationship between learning styles and the acceptance of open learner models for information sharing in technology-based collaborative learning. The result reveals that learning styles have no significant impact on learners' acceptance of open learner models for information sharing. The implications of this study can assist open learner models designers to apply appropriate instructional design strategies in developing open learner models applications.

## 1 Introduction

Technology-based collaborative learning has been increasingly becoming popular for improving collaboration among learners in teaching and learning processes (Stahl et al. 2006; Sridharan et al. 2010; Karunasena et al. 2013). The successful implementation of technology-based collaborative learning depends mainly on learners' willingness to share their learning information through the adoption of collaboration technologies. Sharing learning information in technology-based collaborative learning not only encourages learners' reflection towards their learning. It also improves learners' academic performance (Pai et al. 2014). To realize the potential benefit of information sharing in collaborative learning, many higher educational institutions have introduced collaborative learning tools into their teaching and learning processes (Sridharan et al. 2009). A promising application of such tools for promoting learners' engagement in technology-based collaborative learning is the adoption of open learner models (OLM).

An OLM is a learning data visualization and collaboration tool for representing a learner's current level of knowledge and their misconceptions in a specific subject area (Bull and Kay 2010). With the use of an OLM, learners are able to create an interactive and collaborative learning environment in which they can share learning resources, compare with each other's work, and more importantly self-reflect and self-regulate on their own learning (Govaerts et al. 2010). The adoption of an OLM in learning processes can increase the awareness of learners' knowledge in a particular domain. It helps in the development of learners' meta-cognitive skills, such as self- reflection and self-assessment (Bull and Kay 2010). Furthermore, the use of an OLM in web-based educational settings has the potential to enhance the sharing and dissemination of learning information by facilitating inter-personal interaction and collaboration among learners.

Despite the usefulness of an OLM for facilitating the sharing of learning information in the learning process, the utilization of OLM is not encouraging (Chen et al. 2007). Existing studies try to overcome this issue mainly from the technical perspective through proposing new modes of interactions between OLMs and learners and introducing different visual knowledge representation formats (Bull and Kay 2010). Few studies, however, have empirically investigated the impact of individual learning styles on the adoption of OLM for information sharing in technology-based collaborative learning.

The successful implementation of collaboration technologies requires a proper matching between learners' learning styles and instructional approaches (Taylor 2004). Many studies indicate that learners' willingness to adopt a collaboration technology for facilitating the sharing of information in technology-based collaborative learning is influenced by individual learning styles (Taylor 2004; Cheng 2014; Li 2015). A majority of these studies, however, mainly focus on different educational technologies and knowledge sharing behaviour of learners. The investigation of the impact of individual learning styles on the adoption of OLM for information sharing in technology-based collaborative learning has frequently been ignored. There appears to be a dearth of empirical studies that investigate the impact of learning styles on the acceptance of OLM as an information sharing tool in a technology-based collaborative learning environment.

This study aims to investigate the impact of learners' learning styles on their acceptance of OLM for information sharing in technology-based collaborative learning. It tries to find out whether various learners with different learning styles would have different intentions to adopt OLM for information sharing in technology-based collaborative learning. The data is collected from 240 undergraduate students in Malaysia through an online survey. A chi-square test is performed to investigate the impact of learners' learning styles on their acceptance of OLM for information sharing. The result shows that learning styles do not have a significant impact on learners' intention to use OLM as an information sharing tool. The findings of this study can assist instructional instructors to apply appropriate instructional strategies in adopting OLM for improving the performance of technology-based collaborative learning.

The rest of this paper is organized as follows. In section 2, the related literature is reviewed to justify the need for this study. The research design and the research methodology are discussed in section 3. Subsequently, the findings and the contributions of this study are presented in section 4. Finally, in section 5, the paper concludes with a discussion of the limitations of this study that can be addressed by future research.

## 2   Literature Review and Motivation

Technology-based collaborative learning is a social interaction that involves the acquisition and sharing of experience and knowledge in a community of learners and instructors through the support of collaboration technologies (Stahl et al. 2006). It is an effective and efficient technique for improving learners' learning performance (Stahl et al. 2006). The adoption of such a technique allows learners to exchange ideas and share learning information. The successful implementation of technology-based collaborative learning in teaching and learning very much depends on the willingness of learners to share their learning information through the use of collaboration technologies (Stahl et al. 2006).

Information sharing is about the provision and acquisition of information between individuals within an organization or a group (Pai et al. 2014). It requires an active involvement of learners through adopting specific collaboration tools for generating and transferring knowledge, making collaboration an essential and highly valued process (Stahl et al. 2006; Pai et al. 2014). In technology-based collaborative learning, individual learners can learn more efficiently and effectively by sharing learning information (Stahl et al. 2006; Pai et al. 2014). Recognizing the importance of information sharing in technology-based collaborative learning, collaboration tools such as OLM have been introduced.

OLMs are introduced for helping learners to access their learning information such as learning progress, competency, concepts, and misconceptions (Bull and Kay 2010; Shi et al. 2014). The availability of learners' learning information creates an opportunity for the peers or instructors to provide appropriate scaffolding to learners. It helps in providing learners with a learning environment that allows learners to have a chance to compare their own models with those of peers and instructors as well as enable learners to have an equal opportunity to collaborate critical reflection through sharing learning information with peers (Bull and Kay 2010). In OLM-based collaborative learning, learners are able to compete and work together as a team towards achieving a common goal by sharing and disseminating their learning information.

Despite various advantages associated with information sharing through the utilization of OLM, there are many situations where learning information is not shared effectively through the adoption of OLM. One problematic issue seems to be the low engagement rate on the utilization of OLM for sharing learning information in learning processes. To address this issue, many initiatives have been proposed. Chen et al. (2007), for example, portray an OLM to represent animal companions to increase individual's engagement in collaborative learning. Dimitrova et al. (2000) propose an interactive OLM for attracting learners to negotiate with the model in the learning process. Brusilovsky et al. (2004) introduce additional navigation support in their QuizGuide system to encourage learners' involvement in learning. These studies are conducted for improving learners' participation to utilize OLM for sharing their learning information in collaborative learning. A majority of these studies have focused on the technical issues of an OLM such as the type of interaction and the knowledge representation format. Very little research has been carried out in investigating the impact of learners' learning styles on their acceptance of an OLM for facilitating information sharing.

A learning style refers to the way people receive, process, evaluate, understand, and utilise information in learning (Battaglia 2008). Obtaining information about learners' learning styles can help education instructors apply appropriate instructional strategies in technology-based collaborative learning (Becker 2007). Learners are better motivated to learn in a technology-based collaborative learning environment that accommodates their learning characteristics (Battaglia 2008). Becker (2007), for example, asserts that understanding learners' learning preferences is crucial for improving learners' academic performance in web-based learning. By knowing individual's preferred methods for learning, educational instructors are able to apply appropriate instructional strategies for maximizing the learning potential of each learner (Battaglia 2008).

Learners' learning styles play an important role in determining the acceptance of collaborative technologies for facilitating information sharing in technology-based collaborative learning. Taylor (2004), for example, argues that learners' learning styles have a significant effect on the adoption of knowledge management systems for information sharing. Li (2015) shows that there is a difference among learners with different learning styles towards the adoption of Wikis to facilitate the sharing of learning resources in collaborative learning. Cheng (2014) reveals that some learners would have more advantages on the utilization of Second Life in exchanging learning information in technology-based collaborative learning.

Many studies have been conducted to examine the influence of individual learning styles on the adoption of collaboration technologies for information sharing in a technology-based collaborative learning environment. The investigation about the influence of individual learning styles on the utilization of OLM to facilitate the sharing of individual learner's learning information, however, is still limited. Existing studies show that individuals' adoption decisions are influenced by the attributes of the technology itself and the characteristics of each individual (Sun and Jeyaraj 2013). Different information sharing technologies have their own characteristics. As a result, the acceptance of these technologies in relation to learners' learning styles in different contexts would be different (Sun and Jeyaraj 2013). This shows that there is a need to further examine the effects of learners' learning styles on the acceptance of OLM for information sharing in a technology-based collaborative learning environment.

The information about learners' learning styles is collected based on the VARK learning style model (Fleming 2008). The VARK learning style model is suitable to be adopted for this research because it focuses on those aspects of learning that are particularly significant in designing user friendly interface and determining the efficacy of web-mediated learning (Fleming 2008). Furthermore, VARK also distinctly maps the type of learning materials used for different learning preferences. The VARK model is about how learners receive, interpret, and disseminate information (Fleming 2008). In the VARK learning style model, four categories of modes representing learners' learning styles has been identified. These modes are visual (V), aural (A), read/write (R), and kinaesthetic (K).

Visual learners learn more effectively and efficiently when learning materials are presented in a visual form (Fleming 2008). Visual learners are usually dependent learners who prefer to work in a collaborative learning environment (Fleming 2008). This type of learners is more motivated to engage with a web-based learning environment if collaboration tools are available (Franzoni and Assar 2009). This is because these learners are able to perform well in group discussion, sharing information, and group learning

In contrast, auditory learners prefer to work independently with their learning materials presented in an audio and video form (Pamela 2011). They generally can be classified as reflective learners (Battalio 2009). The collaborative learning environment usually does not create any advantages for this type of learners. As a result, the use of collaboration tools to share their learning information in a collaborative learning environment is not encouraging (Battalio 2009).

Kinesthetic learners learn more efficiently through experience by involving the adoption of a hands-on approach to problem solving. They have same characteristics as active learners where they prefer working in social interaction (Battalio 2009). They prefer physical movement activities in their learning environments such as experimenting and practising which involve multi-sensory experiences. Kinesthetic learners would feel demotivated if their learning environments only allow them to listen and watch in a class passively. They prefer a learning environment which can provide an opportunity for them to discuss and exchange information and interact with each other. In term of read/write learners, they learn best through written and spoken words. They are more motivated to engage in learning where the learning materials are in the form of printed text.

## 3   Research Design and Methodology

The purpose of this study is to investigate the impact of learning styles on learners' acceptance of OLM for information sharing. To achieve this objective, the research question is formulated as follows: Is there a significant difference between learners with different learning styles on the acceptance of OLM for information sharing?

To adequately answer this question, a quantitative research design with a scenario-based and web-mediated prototyping tool is employed as shown in Figure 1. Such a design is frequently used in human-computer interaction research for describing the design specifications and the functionality of a prototype information system to create awareness and brief explanations to the learners before the actual system is developed (Carroll 2000; Sek et al. 2014a). This technique is practical in the initial development of an information system where the feedback from potential learners would put into consideration (Carroll 2000; Sek et al. 2014a). Scenario-based and web mediated prototyping is appropriate for the introduction of new information systems to the potential adopter (Carroll 2000, Sek et al. 2015).

In this study, the description of the OLM sharing features such as functionality, interface layout, and system capability are introduced to the participants in the survey. The purpose of this introduction is to create awareness about the OLM capability in facilitating sharing for collaborative learning. In this prototyping design, three collaboration tools including a model representation tool, a model comparison tool, and a model improvement tool have been proposed for facilitating collaborative learning in an OLM environment.

The purpose of the model representation tool is to visualize learners' learning information such as their competency, knowledge level, and misconceptions. The availability of this information in teaching and learning not only creates awareness of learners about their learning progresses, but also attracts learners' attention. Furthermore, this tool provides reflection on the academic performance of learners in terms of the concepts learnt and misconception existent (Sek et al. 2014b).

The availability of model comparison tools in an OLM learning environment is to allow learners to have an opportunity to compare their own learning progress with other learners. This tool can facilitate information sharing among peers and instructors. Learners are able to make comparison related to their own learning progresses with other learners. If their learning does not progress as expected, learners are able to react immediately to improve their performance by putting in more effort (Sek et al. 2014b).

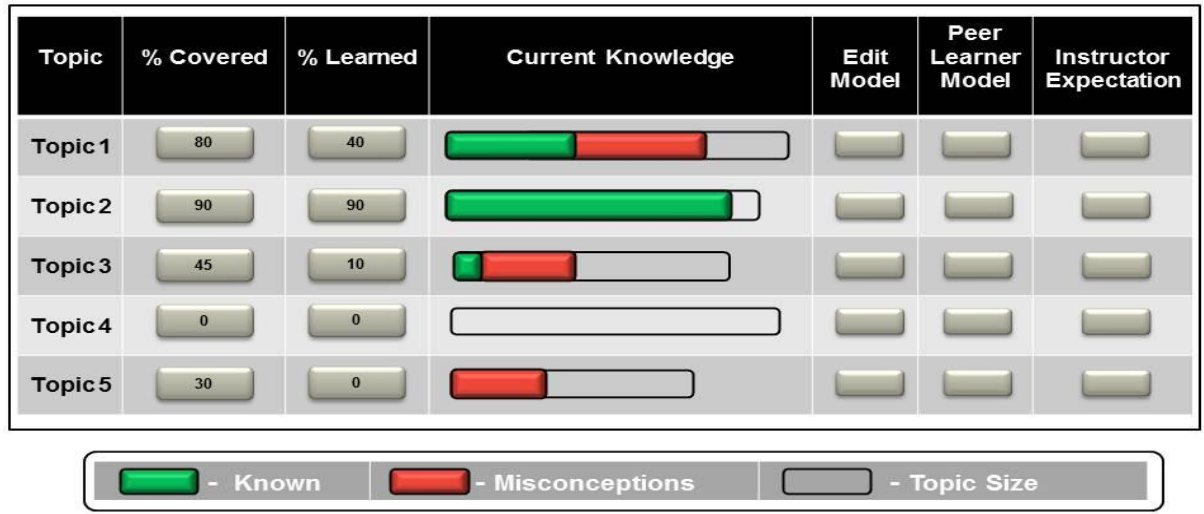

*Figure 1. Web mediated prototyping interface design of OLM*

The model improvement tool provides learners who are lagging behind to have an opportunity to improve their understanding by engaging relevant learning materials provided by the peers and instructors. For example, if learners are able to complete the extra learning materials in the system with a good result, their learning progresses will be updated according to their current levels of understanding. Learners will be supplied with the relevant learning materials by peers or instructors according to their knowledge levels. This is important as it makes learners aware that they have a full control of the outcome of their study. Consequently, this would further improve not only learners' confidence in their learning. It also would increase learners' satisfaction if they are able to accomplish the learning materials provided according to the study plan (Sek et al. 2014b).

Surveys are commonly utilized for gathering information directly from respondents with respect to capturing the views, attitudes and intended behaviours of individual participants on the adoption of specific technologies (Crewell 2003; Sek et al. 2015). This technique is suitable for this study because it employs direct questioning to gather respondents' acceptance on their adoption of an OLM for information sharing. The construction of the survey is based on the previously validated questions designed to reflect an OLM context.

A pilot study is conducted to test and validate the reliability of the questionnaire. The questionnaire consists of three parts. Part-1 involves the participant's profile. Part-2 is about individual learning styles. Part-3 is related to OLM scenarios that are presented to each participant. The focus of these scenarios is to explore an individual learner's acceptance of OLM for sharing learning information.

Convenience sampling is adopted in this study because the selection of the respondents is based on their voluntary and availability (Battaglia 2004; Sek et al. 2015). The participants are undergraduate students, enrolled in a university in Malaysia. This sample is selected because the participants have an experience in engaging learning management systems in their educational coursework. Obtaining the web-based learning experience from these participants provides reliable data regarding their OLM acceptance for information sharing. This study employs a web-mediated survey tool to collect data from the participants. Before assessing the learners' acceptance of OLM for information sharing, the participants are introduced to the OLM through the scenario-based OLM prototype by using Adobe Captivate 7 (Sek et al. 2014a). The online survey related to learners' acceptance of OLM for information sharing is distributed to 260 undergraduate students in a university in Malaysia. A total of 240 participants or approximately 92% has responded to the online survey.

The chi-square test is adopted to examine the relationship between learners' learning styles and their acceptance of OLM for information sharing (Pallant 2010). The applicability of the chi-square test is due to its ability to determine whether there is a significant association between learning styles and learners' acceptance of OLM for information sharing in technology-based collaborative learning.

Information sharing intention refers as a learner's willingness to adopt OLM for information sharing in a collaborative learning environment. The scale to measure learners' intention to share their learning information through the use of OLM is adopted from Chen et al. (2009) with some modification for accommodating the OLM context. The scale to measure learners' intention to share their learning information through the use of OLM is adopted from Chen et al. (2009) with some modifications. Chen et al.'s (2009) scale is suitable for this research because their study focuses on learners' knowledge sharing behavioural in virtual learning environment which is quite similar to an OLM-based collaborative learning environment. A five-point Likert scale is employed ranging from one describing strongly disagree until five to indicate strongly agree for measuring learners' information sharing intention in OLM context.

## 4  Findings and Discussion

This research aims to investigate the impact of individual learners' learning styles on their acceptance of OLM for facilitating information sharing in their teaching and learning processes. Table 1 shows the demographic profile of the respondents in this study. There are 240 participants consisting of 128 male and 112 female. The respondents are from four faculties including engineering, computer science, business, and social science. The purpose of gathering the data from four different faculties is to have respondents with a variety of backgrounds and skills represented.

| Variables | | Frequency | Percentages |
| --- | --- | --- | --- |
| Gender | Male | 128 | 53.0 |
|  | Female | 112 | 47.0 |
| Faculty | Engineering | 57 | 23.7 |
|  | Computer science/ IT | 64 | 26.7 |
|  | Business/Management | 60 | 25.0 |
|  | Social science | 59 | 24.6 |

*Table 1. The Demographic Profile of Respondents*

In this study, there are four learning styles that have been captured based on the VARK learning styles inventory from http://vark-learn.com/the-vark-questionnaire/the-vark-questionnaire-for-younger-people/. Table 2 shows the distribution of these four learning styles and the corresponding mean and standard deviation on the learners' acceptance of OLM for information sharing. From a total of 240 respondents, about 30.8 per cent of the learners are on the aural learning style preference. The visual, read/write, and kinaesthetic learning style preferences appear to share similar percentages of 23.7 per cent, 20.8 per cent, and 24.7 per cent respectively. As indicated in Table 2, the Kinesthetic type of learners has a highest mean score of 3.68 on the acceptance of OLM for information sharing. The second highest mean score is the visual learner which carries a value of 3.63. The read/write type of learners obtained the lowest mean score of 3.18 in accepting OLM for information sharing.

| Learning styles | Frequency | Percentages | Mean | Standard Deviation |
| --- | --- | --- | --- | --- |
| Visual | 57 | 23.7 | 3.63 | 1.23 |
| Aural | 74 | 30.8 | 3.49 | 1.31 |
| Read/Write | 50 | 20.8 | 3.18 | 1.29 |
| Kinesthetic | 59 | 24.7 | 3.68 | 1.32 |

*Table 2. Descriptive Statistics of the Acceptance of OLM for Information Sharing with Different Learning Styles*

Table 3 shows the relationship between learners' learning styles and the level of agreement with the statement "I am willing to collaborate with peers by sharing my learning information using open learner model". It shows that there are no huge variations in the level of agreement among different learning styles on the acceptance of OLM for information sharing. Interestingly, all participants have a similar preference in accepting the adoption of OLM for information sharing in collaborative learning. This implies that OLM provides similar benefits to all types of learners, irrespective of their preferred learning styles. This assumption is confirmed with the results of the chi-square test, $\chi^2 = (12, n = 240) = 10.481$, $p = 0.05$, indicating that significant differences do not exist among different levels of agreement on the acceptance of an OLM for facilitating the sharing of learning information.

|  |  | Learning Styles | | | | |
|---|---|---|---|---|---|---|
|  |  | Visual | Aural | Read/Write | Kinesthetic | Total |
| SDA | Count | 5 | 8 | 8 | 6 | 27 |
|  | % within learning styles | 8.8 | 10.8 | 16.0 | 10.2 | 11.3 |
| DA | Count | 7 | 12 | 7 | 8 | 34 |
|  | % within learning styles | 12.3 | 16.2 | 14.0 | 13.6 | 14.2 |
| N | Count | 6 | 7 | 9 | 3 | 25 |
|  | % within learning styles | 10.5 | 9.5 | 18.0 | 5.1 | 10.4 |
| AG | Count | 25 | 30 | 20 | 24 | 99 |
|  | % within learning styles | 43.9 | 40.5 | 40.0 | 40.7 | 41.3 |
| SAG | Count | 14 | 17 | 6 | 18 | 55 |
|  | % within learning styles | 24.6 | 23.0 | 12.0 | 30.5 | 22.9 |
| Total | Count | 57 | 74 | 50 | 59 | 240 |
|  | % within learning styles | 100 | 100 | 100 | 100 | 100 |

SDA=Strongly Disagree, DA = Disagree, Neutral = N, AG = Agree, SAG = Strongly Agree

*Table 3. The Relationship between Learners' Learning Styles and the Acceptance of OLM for Information Sharing*

Reflective/auditory learners, generally, are not motivated to engage collaboration tools in a collaborative learning environment (Li 2015). They prefer to work independently (Pamela 2011; Battalio 2009). However, the finding as shown in Table 3 indicates that about 64 per cent of the total aural learners intend to accept OLM for sharing their learning information. As indicated in Table 2, the mean score of this type of learners is about 3.49 which showing learners to have an intention to share their learning information through the adoption of OLM. This finding is consistent with that of Battalio (2009) who asserts that reflective learners appear to be more positive in dealing with collaborative learning. Reflective learners value collaborative learning if they are able to work independently when sharing their learning information (Ke and Carr-Chellman 2006). The use of OLM for information sharing allows learners to share their learning information independently and freely without having any difficulties have positively influence the reflective learners towards the adoption of OLM.

On the other hand, active/kinesthetics learners naturally prefer to interact and collaborate more with peers for obtaining learning information. In this study, about 71 per cent of the kinesthetics learners as shown in Table 3 have expressed a preference for adopting OLM for sharing their learning information. The same indication also appear in Table 2 which shows the highest mean score of 3.68 among other types of learners in their willingness to adopt OLM for information sharing. Consistent with Li (2015), this finding lends some support to existing research indicating that kinesthetics learners demonstrate a preference for engaging in a technology-based collaborative learning environment. This shows that the availability of OLM to encourage interaction and collaboration activities has attracted kinesthetics learners to participate in sharing their learning information.

Visual learners are usually dependent learners. They prefer to work in a technology-based collaborative learning environment (Pamela 2011). In Table 3, about 68 per cent of the visual learners have a preference to adopt OLM for sharing learning information. The agreements to the statement also indicating in Table 2 which showing about 3.63 mean score in willingness to use OLM for information sharing. This finding is supported in the research conducted by Franzoni and Assar (2009) showing that visual learners are keen to engage collaboration tools to share their information in technology-based collaborative learning. The features and functionalities available in an OLM learning environment which allows sharing and interaction activities among peers have stimulated learners' interest to engage in this tool.

Read/write learners prefer to participate in a web-based learning environment when learning information is printed in words on the web interface which requires a lot of reading (Battalio 2009). Table 3 shows more than 50 per cent of read/write learners are willing to adopt OLM for information sharing. The evidence also shows in Table 2 which indicating the read/write learning style preference possess a mean score of 3.18 in accepting OLM for information sharing. This is consistent with Battalio (2007) who claims that verbal learners have a positive reaction in the technology-based collaborative learning environment. The interface of an OLM consists of a lot of written reading materials and visual components. The combination of these components has created a suitable learning environment for verbal learners to interact with peers for collaboration learning Battalio (2007).

The contributions of this study to the literature are twofold. Firstly, the study provides a better understanding of the impact of learners' learning styles on their adoption of an OLM for information sharing. Secondly, this study provides insights towards the adoption of an OLM for information sharing from the perspective of different learners' learning styles. Such insights can assist OLM designers to apply appropriate instructional design strategies in developing OLM applications in future. The research findings have practical implications for educational institutions for developing efficient and effective strategies and policies that promote the OLM adoption to facilitate information sharing among learners in technology-based collaborative learning.

## 5 Conclusions, Limitations and Future Research

This study examines the influence of learners' learning styles on their acceptance of an OLM for information sharing in technology-based collaborative learning. The findings show that learners' adoption of an OLM for information sharing is not influenced by their learning styles. The results of this study provide educational instructors with a profound insight into the issues involved in the acceptance of OLM for information sharing in technology-based collaborative learning.

There are some limitations in this study. Firstly, this study only selects one university in Malaysia to investigate the influence of learners' learning styles on their acceptance on the adoption of OLM for information sharing. In order to gain reliable and general view of this acceptance, the selection of the samples can be extended to more universities in Malaysia. Secondly, this study employed scenario-based web-mediated prototyping design for investigating the impact of learners' learning styles on their acceptance towards the adoption of OLM for information sharing. Future study can be conducted to investigate this impact by developing a real OLM learning environment which learners are able to have a real experience in engaging with OLM tools.

## Acknowledgements

This research is supported by the Ministry of Education of Malaysia (MOE) and Universiti Teknikal Malaysia Melaka (UTeM)

## Copyright